\documentclass[12pt]{article}

\title{\bf
                Remote control system of\\
                a binary tree of switches --\\
                I. constraints and inequalities
}

\author{
  O.~Golinelli\\
  \normalsize\itshape Institut de Physique Th\'{e}orique,
  \normalsize\itshape CEA, CNRS, Universit\'{e} Paris-Saclay, France
}

\date{}

\usepackage[a4paper, left=30mm, right=30mm, top=30mm, bottom=34mm]{geometry}

\usepackage[final]{graphicx}

\usepackage[final, breaklinks, colorlinks, allcolors=black]{hyperref}

\begin{document}

\maketitle

\begin{abstract}
  \normalsize  
  We study a tree coloring model introduced by Guidon (2018), initially
  based on an analogy with a remote control system of a rail yard, seen as
  a switch tree.  For a given rooted tree, we formalize the constraints on
  the coloring, in particular on the minimum number of colors, and on the
  distribution of the nodes among colors.  We show that the sequence
  $(a_1,a_2,a_3,\cdots)$, where $a_i$ denotes the number of nodes with
  color $i$, satisfies a set of inequalities which only involve the
  sequence $(n_0,n_1,n_2,\cdots)$ where $n_i$ denotes the number of nodes
  with height $i$.  By coloring the nodes according to their depth, we
  deduce that these inequalities also apply to the sequence
  $(d_0,d_1,d_2,\cdots)$ where $d_i$ denotes the number of nodes with depth
  $i$.
\end{abstract}

\section{Introduction}

Since the appearance of computers, the notion of tree is central in
computer sciences.  For example to decompose program lines, the first
Fortran compilers used a binary tree optimized for keyword recognition
time.  Tree structures are intensively used for databases, algorithms,
representation of expressions in symbolic programming languages, etc.

Many variants have been developed to minimize searching time or to save
memory, like B-tree and red–black tree.  Trees are essential for network
design and parallel computing: many computer clusters have a fat-tree
network and algorithm performances depend on lengths of communication links
between computer nodes.

The mathematician and computer scientist D.~ Knuth, also creator of the TeX
typesetting system, devotes a hundred pages to trees in his encyclopedia,
\emph{The art of computer programming}, and summarizes their development
since 1847~\cite[page 406]{Knuth}.

Moreover coloring problems are frequent in graph theory.  The most common
rule for coloring a graph is that two adjacent vertices do not have the
same color.  Unfortunately this rule has little interest for a tree, which
is a graph without loop: just alternate two colors along each branch.

By leafing through a general public review on Linux, we read an
article~\cite{Guidon-1} that describes how to control a binary tree of
electronic switches with a minimum number of signals.  The
author Y.~Guidon explains how to balance the signal power to avoid the bad
situation where a single signal controls half of the switches.  For that, he
describes this problem in terms of binary tree coloring, but with a rule
different from that usual in graph theory.  He draws balanced
solutions~\cite{Guidon-1, Guidon-2} for perfect binary trees with height up
to 5.

Our paper is devoted to this kind of coloring.  In
Section~\ref{sec:definitions}, we recall some usual definitions on trees.
As Ref.~\cite{Guidon-1} is in French and difficult to access for
non-subscribers, we recall its analogy with a rail yard in
Section~\ref{sec:analogy} and its coloring rule in Section~\ref{sec:color}.
In Section~\ref{sec:partitions}, we study the constraints to distribute the
nodes of a given tree among the colors, according to their distribution by
height in the tree.  In Section~\ref{sec:proof}, we give proofs of the
inequalities stated on Section~\ref{sec:partitions}.

\section{Definitions}

\label{sec:definitions}

To fix terminology, we quickly give some definitions.  Most of them are
common, but some differ according to the authors, like binary, full binary
and perfect binary trees.

A \emph{graph} is defined by a set of \emph{nodes} (or vertices) and a set
of \emph{edges} (or links), where an edge is a pair of nodes.
A \emph{tree} is a connected graph without cycle.
The \emph{size} of a tree is the number of its nodes.

A \emph{rooted tree} is a tree with a marked node, called the \emph{root};
edges are now oriented away from the root.  A rooted tree can be described
as a family chart, with a common ancestor and the descendants.  By
convention, the root is drawn at the top, unlike real trees in nature.

The \emph{depth} of a node $v$ is the number of edges between $v$ and the
root.  Only the root has depth $0$.  A node with depth $d$ can share an
edge with a \emph{child}, a node with depth $d+1$, or a \emph{parent},
a node with depth $d-1$.

Each node has only one parent, except the root without parent.  Considering
the number of children, there are two kinds of nodes: \emph{leaves} with no
children, and \emph{internal nodes} with at least one child.

A \emph{sibling} of a node $v$ is a node $w$ that has same parent as $v$,
with $w \ne v$.
A \emph{descendant} is a child, or a descendant of a child;
an \emph{ancestor} is a parent, or an ancestor of a parent (recursive
definitions).

The \emph{height} of a node $v$ is the maximal distance between $v$ and the
leaves among its descendants; leaves have height 0.  The height of an
internal node is $1 + \max(\{h_c\})$, where the $h_c$'s are the heights of
its children.

The \emph{height of a rooted tree} is the height of its root; it is also
the greatest depth of its leaves.

\begin{figure}
  \centering
  \includegraphics{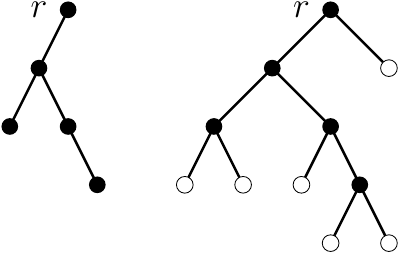}
  \caption{Left: a binary tree with root $r$ and $n=5$ nodes.  Right: a
    full binary tree with $2n+1=11$ nodes ($n=5$ internal nodes in black
    and $n+1=6$ leaves in white).  These two trees are in bijection by
    adding or removing white nodes; see text.}
  \label{fig:bijection}
\end{figure}

A \emph{binary tree} is a kind of rooted tree where each node has at most 2
children, distinguishing \emph{left child} and \emph{right child}, even
when there is only one child.  In other words, each node has 0 or 1 left
child, and 0 or 1 right child.  For example, see
Fig.~\ref{fig:bijection}.

A \emph{full binary tree} is a binary tree where each node has 0 or 2
children.  For example, see Fig.~\ref{fig:bijection}.  A full binary tree
with $n$ internal nodes has $n+1$ leaves, so $2n+1$ nodes.

\label{sec:bijection}
There is a bijection \cite[page 16]{Stanley-book} between the set
$\mathcal{F}_n$ of full binary trees with $2n+1$ nodes and the set
$\mathcal{B}_n$ of binary trees with $n$ nodes.
Let $T_F \in \mathcal{F}_n$; this bijection consists in removing all the
$n+1$ leaves of $T_F$ and keeping the skeleton made of the $n$ internal
nodes: It is a binary tree $T_B \in \mathcal{B}_n$;
see Fig.~\ref{fig:bijection}.

Starting with $T_B \in \mathcal{B}_n$, the reverse bijection consists in
completing with leaves each node of $T_B$ that has less than two children.
The new tree $T_F \in \mathcal{F}_n$ and its $n$ internal nodes are the
nodes of $T_B$.

Consequently there are as many trees in $\mathcal{F}_n$ as in
$\mathcal{B}_n$.  It is a classical exercise to show that they are counted
by Catalan numbers, $C_n = (2n)! / (n+1)! / n!$, sequence A000108 in
OEIS~\cite{OEIS}.  A book~\cite{Stanley-book} of R.~P.~Stanley presents 212
other kinds of object that are counted using Catalan numbers.

A \emph{perfect binary tree} is a full binary tree in which all the leaves
have same depth.  All the perfect binary trees with a given height $h$ are
isomorphic: Such a tree has $2^d$ nodes with depth $d$ and height $h-d$ for
$0 \le d \le h$, then $2^h-1$ internal nodes and $2^h$ leaves, so
$2^{h+1}-1$ nodes.  By removing the leaves, the sub-tree made of its
internal nodes is a perfect binary tree with height $h-1$.
Fig.~\ref{fig:pbt} represents the perfect binary tree with height 2.

Among all the (full or not) binary trees, perfect binary trees are the most
compact: Given a height $h$, a binary but not perfect tree has less
than $2^{h+1}-1$ nodes.

\section{Analogy with a rail yard}

\label{sec:analogy}

In this section, we follow the analogy made by
Guidon~\cite{Guidon-1,Guidon-2}, between a full binary tree and a rail
yard.  Internal nodes are \emph{switches}.  Cars enter the tree through the
root which is the first switch; then they are directed to leaves, which are
exit tracks.

In a full binary tree, each switch has one input (parent) and two outputs
(children).  Each switch can be oriented to the right or left.  The purpose
of a rail yard is to choose an exit track, i.e. a leaf $v$, and to bring a
car from the root $r$ to $v$.  For that, it is necessary to orient
correctly the switches along the path from $r$ to $v$, i.e. the ancestors
of $v$.  Generalization to all rooted trees is possible, by using switches
with a variable number of outputs.

To easily operate a rail yard, it is convenient to control the switches
remotely: We can send a binary signal $b_u$ to each switch $u$, for example
$b_u=0$ for left or 1 for right.  Consequently for a leaf $v$ with depth
$d$, we need to send $d$ signals to the $d$ switches which are the
ancestors of $v$.  But we must be able to select all leaves, then to
control all switches.  Notice that leaves are passive and do not receive
signal.

For a rail yard made up of $n$ switches, the maximum solution
\label{sec:maxsol} is to have a remote control system with $n$ signals,
where each signal activates one and only one switch.  In contrast,
Guidon~\cite{Guidon-1} asks the question of decreasing or even minimizing
the number of signals.  For practical examples (rail yard, electronic
circuit, etc.), this may have an economic justification, to simplify the
control system and reduce its cost.

For that, some signals must control more than one switch.  But there are
constraints: For each leaf $v$ with depth $d$, its $d$ ancestors must be
controlled by $d$ independent signals.  Since these signals can only take
two values, left=0 or right=1, there will be pairs $\{u, w\}$ of ancestors
of $v$ with signals of the same value, $b_u=b_w$.  However, these two
signals must remain independent, to be able to access other leaves,
especially those that also have $u$ and $w$ as ancestors but with $b_u \ne
b_w$.  In other words, the same signal can control several switches
provided there are no ancestor-descendant pairs among these switches.

\section{Colored rooted trees}

\label{sec:color}

\subsection{Coloring rule}

From now on, we use coloring terminology by assigning a different label, called
\emph{color}, for each signal~\cite{Guidon-1}.  The problem now is to color
all the internal nodes of a full binary tree with the following constraint:
\emph{For each leaf, its ancestors must all be of different colors.}

Note that the leaves are not colored; so we remove them.  As explained in
Section~\ref{sec:bijection}, this operation is a bijection that transforms a
\emph{full binary} tree $T_F$ with $n$ colored internal nodes into a
\emph{binary} tree $T_B$ with $n$ colored nodes (internal or leaves).

By definition, each node of a binary tree has at most two children,
distinguishing left child and right child.  This distinction is important
in the analogy with a rail yard, but we can ignore it when we formulate the
problem as a colored tree.  We can also remove the constraint of having at
most two children per node.  So we can therefore generalize to all rooted
trees.

Our problem is now reformulated as follows: color all the nodes of a rooted
tree with the following rule:
\emph{For each leaf $v$, $v$ and its ancestors must all be of different color.}
This is equivalent to the:

\medskip\noindent\textbf{Coloring rule:}
\label{sec:rule}
\emph{For each pair of nodes $\{u,v\}$, if $u$ is an ancestor of $v$, they
      do not have the same color.}
\medskip

We remark that this constraint is stronger than the usual rule for graph
coloring, such that two connected nodes do not have the same color.

As for graph coloring theory, values of color label do not matter.  We can
change or permute the labels without effects.  Only counts the partition of
the tree into subsets of nodes having the same color: Two colorings are
equivalent if they give the same subsets of nodes.

\subsection{Minimum number of colors}

For a rooted tree of $n$ nodes, the maximum solution indicated above
Section~\ref{sec:maxsol} corresponds to the \emph{maximum} number $n$ of
colors: one color per node.  But we are rather interested in $\chi$, the
\emph{minimum} number of colors, which would be the equivalent of chromatic
number in graph theory.

\begin{figure}
  \centering
  \includegraphics{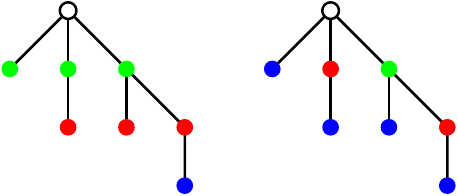}
  \caption{Left: Canonical coloring by depth.  Right: Canonical coloring by
           height with blue, red, green and white for nodes with
           height 0, 1, 2 and 3.}
  \label{fig:canonical}
\end{figure}

For a rooted tree with height $h$, Guidon~\cite{Guidon-1} shows that
$\chi=h+1$.  Indeed according to the definition of $h$, the maximum depth
of a leaf is $h$; therefore $h+1$ different colors are already required for
a leaf with depth $h$ and its $h$ ancestors.  Consequently $\chi \ge h+1$.
To prove equality, it remains to show a solution with $h+1$ colors.  We
will give two examples of minimum solutions in Fig.~\ref{fig:canonical}.

\label{sec:canonical}
The first minimum solution is the \emph{canonical coloring by depth}, where
all nodes with depth $d$ have the same color $d+1$, for $0 \le d \le h$.
The second one is the \emph{canonical coloring by height}, where all
nodes with height $g$ have the same color $g+1$, for $0 \le g \le h$.

Note that for a perfect binary tree, both canonical colorings, by depth or
by height, are equivalent; indeed, all the nodes with depth $d$ have the
same height, $g=h-d$.

\subsection{Balancing of the nodes among colors}

Guidon~\cite{Guidon-1} asks a more difficult question: For a given tree,
how to best balance the nodes among all the colors?  Again the
justification can be economical: If a signal controls many switches (or
electrical relays), its circuit must have a high pneumatic (or electrical)
power.  It may also be slow because of the inertia of the switches.  Also,
we may be interested in a more balanced distribution to minimize total
cost, or reaction times.

Let $T$ be a rooted tree with $n$ nodes, height $h$, and colored with the
minimum number $\chi = h+1$ of colors indexed by $i=1,2,..,\chi$.  Let
$a_i$ be the number of nodes with color $i$.  In our problem, $T$ is fixed:
Work optimization relates only to $A = (a_1, a_2, \cdots, a_\chi)$,
i.e. the way to color $T$, but without modifying edges of $T$.

\begin{figure}
  \centering
  \includegraphics{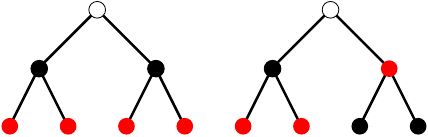}
  \caption{The two possible colorings with 3 colors for the perfect binary
           tree of height 2.}
  \label{fig:pbt}
\end{figure}
For example, for a perfect binary tree with height $h$ and $2^{h+1}-1$ nodes,
the canonical coloring (by depth or equivalently by height) gives the
distribution $A=(2^h, \cdots, 4, 2, 1)$ where more than half of the nodes
have the same color.  We will see that it is the maximally unbalanced
coloring.  But we can color it differently.  For example, for a perfect
binary tree with height 2, there are two possible colorings with $\chi=3$
colors, drawn in Fig.~\ref{fig:pbt}: the canonical coloring $A=(4,2,1)$ and
another coloring $A'=(3,3,1)$, better balanced~\cite{Guidon-1}.

For perfect binary trees, Guidon~\cite{Guidon-1} offers a solution and an
algorithm to minimize $\max_i(a_i)$, equivalent to the maximum power of
control signals.  But we could also choose to minimize another quantity,
for example, $\sum_i a_i^2$, or more generally the $p$'th moment
$\mu_p = \sum_i a_i^p$.  Note that the maximum corresponds to the limit
$p \to \infty$.  We may also want to maximize the coloring entropy
$S = - \sum_i a_i \log(a_i)$, knowing that
$S = - \left. d \mu_p/dp \right|_{p=1}$.  If we know the cost function
$f(a_i)$ of a circuit which controls $a_i$ switches, we will try to
minimize the total cost $\sum_i f(a_i)$.

\section{Colorable partitions}

\label{sec:partitions}

We say that the non-negative integer sequence $A = (a_1, a_2,...,a_c)$ is a
\emph{colorable partition} of a rooted tree $T$ if there exists a coloring
of $T$ with $c$ colors where $a_i$ is the number of nodes of $T$ with color
$i$ for $1 \le i \le c$.  In this section, we will study the conditions
that the coloring partitions must satisfy.

To simplify the proof presented Section~\ref{sec:proof}, we allow more
colors than the minimum, so $c \ge \chi = h+1$ where $h$ denotes the height
of $T$.  Moreover we allow colors which do not color any nodes, i.e. with
$a_i=0$.  Note that this allows to have a number $c=\infty$ of colors, then
$c$ is no longer a relevant parameter.

\subsection{Necessary conditions}

Let $T$ be a rooted tree with $n$ nodes, $h$ the height of $T$, and $n_g$
the number of nodes of $T$ with height $g$ for $g \ge 0$.
If $A = (a_1, a_2,...,a_c)$ is a colorable partition of $T$, then $A$ must
satisfy the following conditions:
\begin{eqnarray}
    && \sum_{i=1}^c a_i= n , \label{eq:sum}\\
    && a_i = 1 \ \ \ \ \mbox{ for at least one color } i \ \ (1 \le i \le c),
            \label{eq:a1}\\
    && a_i \le n_0 \ \ \ \ (1 \le i \le c), \label{eq:ineq_1}\\
    && a_i+a_j \le n_0 + n_1 \ \ \ \ (1 \le i < j \le c), \label{eq:ineq_2}
\end{eqnarray}
and more generally, for any subset of $k$ different colors
$1 \le i_1 < i_2 < \cdots < i_k \le c$  with $1 \le k \le c$,
\begin{equation}
    a_{i_1} + a_{i_2} \cdots + a_{i_k} \le n_0+n_1 \cdots + n_{k-1} .
    \label{eq:ineq_k}
\end{equation}

Inequalities~(\ref{eq:ineq_1}--\ref{eq:ineq_k}) are equivalent to
\begin{eqnarray}
    \max_{1 \le i \le c}(a_i)        & \le & n_0 ,      \label{eq:max_1}\\
    \max_{1 \le i < j \le c}(a_i+a_j) & \le & n_0+n_1 ,  \label{eq:max_2}
\end{eqnarray}
and more generally, for $1 \le k \le c$,
\begin{equation}
    \max_{1 \le i_1 < i_2 < \cdots < i_k \le c}( a_{i_1} + a_{i_2} \cdots + a_{i_k})
    \le n_0+n_1 \cdots + n_{k-1} .  \label{eq:max_k}
\end{equation}

In general, these are \emph{necessary} but \emph{not sufficient}
conditions.  In Section~\ref{sec:tnsc}, we will give examples of rooted,
binary and full binary trees with sequences which satisfy these conditions
but which do not correspond to any possible coloring.

Before proving these conditions in Section~\ref{sec:proof}, we discuss
first some of their consequences.

\subsection{Canonical coloring by height}

We first note that the sequence $(n_0, n_1 \cdots)$ satisfies
$\sum_i n_i =n$, the total number of nodes; it is also monotonically
decreasing (non-increasing).  Indeed, for every $i\ge 0$, each internal
node with height $i+1$ has at least one child with height $i$.  Moreover
each node with height $i$ (except the root) has a single parent, which can
be height $i+1$ or greater.  Consequently $n_i \ge n_{i+1}$.  Note that
$n_i=0$ if $i>h$, because $h$ is the maximum height of nodes of $T$, by
definition.

In Section~\ref{sec:canonical}, we described the canonical coloring by
height where all the nodes with the same height have the same color.  So
the sequence $(n_0,n_1,n_2,\cdots)$ is therefore a colorable partition of
$T$: it satisfies the above conditions with $a_i=n_{i-1}$.  But here
moreover, inequalities~(\ref{eq:max_1}--\ref{eq:max_k}) become
\emph{equalities}.

As the $n_i$'s are monotonically decreasing, the colorable partition
$(n_0,n_1,n_2,\cdots)$ is the largest in lexicographical order among all
the colorable partitions of $T$.  In this sense, we can say that the
canonical coloring by height is maximally unbalanced.

The inequality~(\ref{eq:max_1}) gives a procedure to maximize the use of a
color: color all leaves of $T$ with this color.

If we want to maximize $k$ colors one after the other, just color with the
color $i$ the nodes with height $i-1$ for $1 \le i \le k$.

To maximize $k$ colors overall, following inequality~(\ref{eq:max_k}), just
color with these $k$ colors all the nodes of the first $k$ levels, i.e. the
nodes with height $i<k$; we can mix colors by levels, according to the
different branches of the tree, but always respecting the coloring rule.

\subsection{Canonical coloring by depth}

Thanks to the inequalities~(\ref{eq:ineq_1}--\ref{eq:max_k}), we obtain a
non trivial relation between the distribution of the nodes by
\emph{depth} and their distribution by \emph{height}.

Let $d_i$ be the number of nodes of $T$ with depth $i$.  In
Section~\ref{sec:canonical}, we described the canonical coloring by depth
where all the nodes with the same depth have the same color.
So $D = (d_0,d_1,d_2,\cdots)$ is a colorable partition of $T$.  Generally,
the sequence $D$ is not necessarily increasing or decreasing.  However, $D$
satisfies the inequalities~(\ref{eq:ineq_1}--\ref{eq:max_k}) with
$a_i=d_{i-1}$.

For example, the tree displayed in Fig.~\ref{fig:canonical} has a
distribution by depth $D=(1,3,3,1)$ and a distribution by height
$(n_0,n_1,n_2,n_3) = (4,2,1,1)$.  For this tree,
Eq.~(\ref{eq:max_1}) gives $3\le 4$,
Eq.~(\ref{eq:max_2}) gives $3+3 \le 4+2$,
Eq.~(\ref{eq:max_k}) gives $3+3+1 \le 4+2+1$ for $k=3$,
and $3+3+1+1 \le 4+2+1+1$ for $k=4$.

\subsection{Necessary but not sufficient conditions}

\label{sec:tnsc}

We prove in Section~\ref{sec:proof} that Eqs.~(\ref{eq:sum}--\ref{eq:ineq_k})
are necessary conditions, in other words if $A$ is a colorable partition of
a rooted tree, then $A$ satisfies these equations.

We can easily draw trees for which these conditions are
\emph{necessary and sufficient}, but this is not generally true.  There are
trees $T$ for which these conditions are not \emph{sufficient}, i.e.  there
is at least one sequence $A=(a_1,a_2,a_3,\cdots)$ which satisfies these
conditions but which do not correspond to any possible coloring of $T$ such
that $a_i$ counts the nodes colored by $i$.  We will give some examples of
trees with non-sufficient conditions (TNSC).

\begin{figure}
  \centering
  \includegraphics{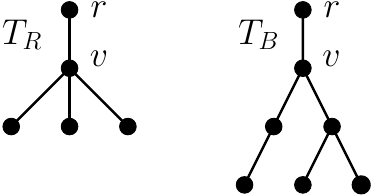}
  \caption{The smallest rooted tree (left) and binary tree (right) with
           non-sufficient conditions.}
  \label{fig:trtb}
\end{figure}
The tree $T_R$ drawn in Fig.~\ref{fig:trtb} is the smallest TNSC among
rooted trees.  It has $n=5$, $h=2$, and $(n_0,n_1,n_2)=(3,1,1)$.
The sequence $A=(2,2,1)$  satisfies all the conditions, because
$2 \le 3$, $2+2 \le 3+1$, and $2+2+1 \le 3+1+1$.

However $A$ is not a colorable partition of $T_R$.  Indeed the root $r$ has
only one child $v$.  Let $i$ be the color of $r$, and $j$ the color of $v$.
The node $v$ is the only node with the color $j$ because all other
nodes of $T_R$ are either ancestor or descendant of $v$.  So we must have
at least two colors with $a_i=a_j=1$, which is not possible with $A=(2,2,1)$.

For binary trees, the smallest TNSC is $T_B$ (up to isomorphism) drawn
in Fig.~\ref{fig:trtb}, for which $n=7$, $h=3$ and
$(n_0,n_1,n_2,n_3)=(3,2,1,1)$.
The sequence $A=(2,2,2,1)$ satisfies all the conditions, but $A$ is not a
colorable partition of $T_B$.  Again, like for $T_R$, the root $r$ has only
one child $v$; we must have at least two colors with $a_i=a_j=1$, which is
not possible.  We can verify that $T_B$ is the only TNSC among binary trees
with size $n \le 7$.  On the other hand, it is easy to draw binary TNSC
with size $n \ge 8$.

With these examples, we see that we can improve Eq.~(\ref{eq:a1}) with the
condition that if $T$ has a unique path from the root to the depth $d$,
i.e. $T$ has only one node with depth $g$ for $0 \le g \le d$, then we have
$a_i=1$ for at least $d+1$ different colors.  But despite this additional
necessary condition, these are still not sufficient conditions.

\begin{figure}
  \centering
  \includegraphics[width=\columnwidth, height=47mm, keepaspectratio]{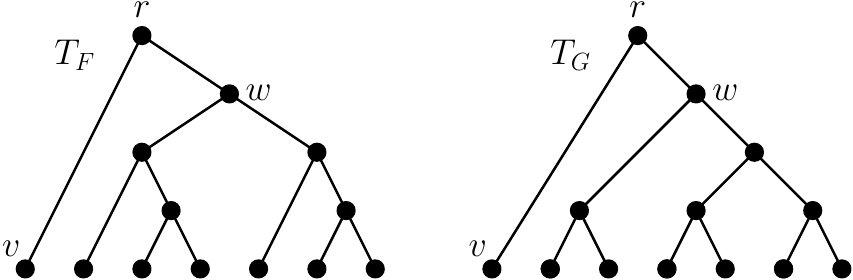}
  \caption{The two smallest full binary trees  with non-sufficient conditions.}
  \label{fig:tftg}
\end{figure}
Indeed, the full binary trees are not affected by this situation, but there
are also full binary TNSC.  The two smaller ones are $T_F$ and $T_G$
(up to isomorphism) drawn in Fig.~\ref{fig:tftg}, with $n=13$ nodes, $h=4$ and
respectively $(n_0,\cdots,n_4) = (7,2,2,1,1)$ and $(7,3,1,1,1)$.
For $T_F$ and $T_G$, the sequence $A = (3,3,3,3,1)$ verifies the necessary
conditions, but $A$ is not a colorable partition.

Let $r$ be the root, $v$ the child of $r$ without descendant, and $w$ the
other child of $r$ (see Fig.~\ref{fig:tftg}).  Let $i$ be the color of the root
$r$, and $j$ the color of $w$.  The color $j$ cannot be assigned either to
the ancestor $r$, or to descendants of $w$; only $v$ can possibly be
colored by $j$.  So $a_i=1$ for $r$ and $a_j \le 2 $ for $w$.  It is not
possible with $A = (3,3,3,3,1)$.

We deduce that for every rooted tree $T$, we have a constraint on the color
$j$ of each child $v$ of the root $r$: $j$ can only appear among the
siblings of $v$ or their descendants.  Therefore
\begin{displaymath}
    a_j \le 1 + \sum_{w \in S(v)} n_0(T_w)
\end{displaymath}
where  $S(v)$ denotes the set of siblings of $v$, $T_w$ the subtree of $T$
rooted in the node $w$, and $n_0(t)$ the number of leaves of the subtree $t$.

We can generalize to each node $v$ with depth $d$: Its color $j$ can only
appear among the siblings of $v$ or their descendants, or the siblings of
the $d-1$ ancestors (except the root) of $v$ (i.e. uncles and aunts,
great-uncles and great-aunts, etc) and their own descendants.  Therefore
\begin{equation}
    a_j \le 1 + \sum_{\lambda=0}^{d-1} \sum_{w \in S(p^\lambda(v))} n_0(T_w)
    \label{eq:aj}
\end{equation}
where $p(v)$ denotes the parent of $v$, and $p^\lambda(v)$ the $\lambda$'th
ancestor of $v$, with $p^0(v)=v$.

We could add these new conditions to Eqs.~(\ref{eq:sum}--\ref{eq:ineq_k}).
But these constraints depend on the fine details of the structure of $T$.
In general, they cannot be expressed simply in terms of global data
such as the number of nodes by height, depth, etc.

We see that these conditions give strong constraints when the tree is
unbalanced, like those given as examples.  Indeed for $T_R$ and $T_B$, the
node $v$ has no sibling, $S(v)=\emptyset$ and $d=1$, so Eq.~(\ref{eq:aj})
gives $a_j \le 1$ as written above.  For $T_F$ and $T_G$, $v$ has only
one sibling $w$, with a subtree $T_w$ reduced to only $w$, so $n_0(T_w)=1$:
Eq.~(\ref{eq:aj}) gives $a_j \le 1+1$.

On the contrary, a perfect binary tree is maximally balanced.  Let $T$ be a
perfect binary tree with height $h$ and $n=2^{h+1}-1$ nodes.  For a node
$w$ with height $g$ ($0 \le g \le h$), the subtree $T_w$ has
$n_0(T_w) = 2^g$ leaves.  Consequently for the color $j$ of a node with
height $g$, Eq.~(\ref{eq:aj}) gives $a_j \le M(g)$ where
\begin{displaymath}
    M(g) = 1 + \sum_{p=g}^{h-1}2^p = 2^h - 2^g + 1.
\end{displaymath}
For the root, $g=h$ and $M(h)=1$: We again find the constraint $a_j=1$.
For other nodes, $g<h$ and the constraint is weak because $M(g) \ge 1 +
2^{h-1} > n/4$: each color (except the root color) is allowed for at least
$1/4$ of the nodes.

Also for perfect binary trees, we conjecture that the necessary conditions
Eqs.~(\ref{eq:sum}--\ref{eq:ineq_k}) are also \emph{sufficient} conditions.

\section{Proof of Eqs.~(\ref{eq:sum}--\ref{eq:ineq_k})}

\label{sec:proof}

We will now prove that Eqs.~(\ref{eq:sum}--\ref{eq:ineq_k}) are indeed
necessary conditions for a sequence $(a_1, a_2, \cdots, a_c)$ to be a
colorable partition of a rooted tree $T$.

Eq.~(\ref{eq:sum}) simply says that all the nodes of $T$ are colored.
Eq.~(\ref{eq:a1}) corresponds to the color $i$ of the root $r$.  Since $r$
is the ancestor of all other nodes, $r$ is the only node colored with
$i$, so $a_i=1$.  Note that we can have others colors $j$ with $a_j=1$.

Proving inequality Eqs.~(\ref{eq:ineq_1}--\ref{eq:ineq_k}) is more
complicated.  When $k \ge h+1$, we note that $n_0+n_1 \cdots + n_{k-1} = n$
and Eq.~(\ref{eq:ineq_k}) becomes $a_{i_1} + a_{i_2} \cdots + a_{i_k} \le n$:
it is always true for all different colors $i_{\lambda}$ because of
Eq.~(\ref{eq:sum}).

When $k=1$, Eq.~(\ref{eq:ineq_k}) is reduced to Eq.~(\ref{eq:ineq_1}).  For
a given color $i$, for each leaf $v$ of the rooted tree $T$, the coloring
rule defined in Section~\ref{sec:rule} requires that there is at most one
node $w$ colored by $i$ on the path from the root to $v$.  Conversely for
each node $w$ colored by $i$, there is at least one leaf $v$ where $w$ is
on the path from the root to $v$.  As $n_0$ denotes the number of leaves,
this ensures that $a_i \le n_0$ for all colors $i$.

This reasoning is simple and valid, but it is difficult to generalize for
several colors ($k > 1$).  Also we will give another proof for $k = 1$, but
easily generalizable to $k > 1$.  For this, we use a recursive definition:
a rooted tree
\begin{displaymath}
    T = (r, \{T_1, T_2, ..., T_q\})
\end{displaymath}
consists of a node $r$, the root, and  a (possibly empty) set of rooted trees
$T_f$, called the \emph{subtrees} of $T$.

For a given rooted tree $T$, we denote by $h(T)$ the height, $n(T)$ the
size i.e. the number of nodes, $n_g(T)$ the number of nodes with height
$g$.  For a coloring of $T$, we denote by $a_i(T)$ the number of nodes with
color $i$.  In principle, this coloring respects the coloring rule and
$(a_1(T), a_2(T), \cdots, a_c(T))$ is a colorable partition of $T$.

Let $T_f$ be a subtree of $T$.  By considering $T_f$ as an isolated rooted
tree, the restricted coloring on $T_f$ always respects the coloring rule,
since the restriction on $T_f$ does not give additional constraints.  We
deduce that $(a_1(T_f),a_2(T_f)\cdots, a_c(T_f))$ is a colorable partition
of $T_f$, therefore its must respect Eqs.~(\ref{eq:sum}--\ref{eq:ineq_k}).

With this division of a tree into root and subtrees, we will be able to
give an proof by induction on the height of the trees, because $h(T_f) <
h(T)$.  Before discussing any value of $k$, we will explain in details the
cases $k = 1$ and 2.

\subsection{Proof of Eq.~(\ref{eq:ineq_1})}

We will prove that $a_i(T) \le n_0(T)$ by induction on the height $h(T)$.
The initial case is $h(T)=0$, for which there is only one rooted tree,
$T = (r, \emptyset)$, reduced to its root and without subtrees, with
$n_0(T)=n(T)=1$.  As $a_i(T)=0$ or 1 for any color $i$,
inequality~(\ref{eq:ineq_1}) is always true.

We now consider a colored rooted tree $T$ with $h(T) \ge 1$ and a color
$i$; we assume the induction hypothesis that $a_i(T') \le n_0(T')$ for all
rooted trees $T'$ with $h(T') < h(T)$.  As $h(T) \ge 1$, $T$ has at least
one subtree.  We will distinguish two cases, depending on whether the root
$r$ of $T$ is colored by $i$ or not.

If $r$ is colored by $i$, it is the only node with this color: $a_i(T)=1$.
As the number of leaves $n_0(T) \ge 1$, inequality~(\ref{eq:ineq_1}) is true.

If $r$ is not colored by $i$, then all the nodes colored by $i$ are in the
subtrees $T_f$ of $T$, i.e. $a_i(T) = \sum_f a_i(T_f)$.  The induction
hypothesis applies to each subtree $T_f$ because $h(T_f) < h(T)$; so
$a_i(T_f) \le n_0(T_f)$.  Since all the leaves of $T$ are also the leaves
of its subtrees, $\sum_f n_0(T_f) = n_0(T)$.  Summing on all the subtrees,
we get that $a_i(T) \le n_0(T)$.

By induction on the height, Eq.~(\ref{eq:ineq_1}) is true for all rooted trees.

\subsection{Proof of Eq.~(\ref{eq:ineq_2})}

We note that Eq.~(\ref{eq:ineq_2}) corresponds to Eq.~(\ref{eq:ineq_k})
with $k = 2$.  We will prove it by induction on the height $h(T)$.  As
explained above, Eq.~(\ref{eq:ineq_k}) holds for $k \ge h(T)+1$.
Therefore Eq.~(\ref{eq:ineq_2}) is valid for the rooted trees $T$ with
height $h(T) \le 1$.  This is the initial case of our proof by induction.

We now consider a rooted tree $T$ with $h(T) \ge 2$, and we assume the
induction hypothesis that Eq.~(\ref{eq:ineq_2}) is true for all rooted trees
$T'$ with $h(T') < h(T)$.  We will distinguish two cases, depending on
whether the root $r$ of $T$ is colored either by $i$ or $j$, or by a third
color.

If $r$ is colored by $j$ (or equivalently by $i$), then $a_j(T) = 1$ and
$a_i(T) = \sum_f a_i(T_f)$.  As $a_i(T_f) \le n_0(T_f)$ for each subtree
$T_f$ of $T$ and $\sum_f n_0(T_f) = n_0(T)$, we get
$a_i(T)+a_j(T) \le n_0(T) + 1$.  As $h(T) \ge 2$, $n_1(T) \ge 1$ and
Eq.~(\ref{eq:ineq_2}) is validated.

If $r$ is not colored by $i$ or $j$, then $a_i(T) = \sum_f a_i(T_f)$ and
$a_j(T) = \sum_f a_j(T_f)$.  So
$a_i(T) + a_j(T) = \sum_f (a_i(T_f) + a_j(T_f))$.  The induction hypothesis
applies to each subtree $T_f$, i.e.  $a_i(T_f) + a_j(T_f) \le n_0(T_f) +
n_1(T_f)$.  Summing on all the subtrees, $a_i(T) + a_j(T) \le \sum_f
(n_0(T_f) + n_1(T_f))$.

As the height of the root $r$ is $h(T) \ge 2$, $r$ does not count among the
nodes of height 0 or 1.  So $\sum_f n_0(T_f) = n_0(T)$ and
$\sum_f n_1(T_f) = n_1(T)$.  Consequently Eq.~(\ref{eq:ineq_2}) is true in
all cases and for all heights.

\subsection{Proof of Eq.~(\ref{eq:ineq_k})}

The proof of Eq.~(\ref{eq:ineq_k}) for $k \ge 2$ colors is a generalization
of the case $k = 2$.  This is a double proof by induction, first on $k$,
then on the height $h(T)$.

For $k$, the initial case is $k=1$, i.e. the already proven
Eq.~(\ref{eq:ineq_1}).  We now consider $k \ge 2$ and we assume the
induction hypothesis that Eq.~(\ref{eq:ineq_k}) is true for $k-1$ colors.

As explained above, Eq.~(\ref{eq:ineq_k}) holds when $h(T) \le k-1$.  This
is the initial case for the induction on $h(T)$.  We now consider a rooted
tree $T$ with $h(T) \ge k$, and we assume the induction hypothesis that
Eq.~(\ref{eq:ineq_k}) is true for $k$ colors and for all rooted trees $T'$
with $h(T') < h(T)$.

We will distinguish two cases, depending on whether the root $r$ of $T$ is
colored either by one of the $k$ colors $i_1,i_2,\cdots, i_k$, or by
another color.

If $r$ is colored by $i_k$ (or equivalently by a color $i_\lambda$ with
$\lambda<k$), then $a_{i_k}(T) = 1$ and
$a_{i_\lambda}(T) = \sum_f a_{i_\lambda}(T_f)$ for $1 \le \lambda \le k-1$.
The induction hypothesis applies to each subtree $T_f$ for $k-1$ colors:
\begin{displaymath}
    a_{i_1}(T_f) + \cdots + a_{i_{k-1}}(T_f) \le n_0(T_f) + \cdots + n_{k-2}(T_f).
\end{displaymath}
As the height of the root $r$ is $h(T) \ge k$, $r$ does not count among the
nodes of height $g$ for $g<k$ and $\sum_f n_g(T_f) = n_g(T)$.  Summing on
all the subtrees,
\begin{displaymath}
    a_{i_1}(T)+ \cdots + a_{i_{k-1}}(T) + a_{i_k}(T)
      \le n_0(T)+\cdots + n_{k-2}(T) + 1.
\end{displaymath}
As $n_{k-1}(T) \ge 1$ because $h(T) \ge k$, Eq.~(\ref{eq:ineq_k}) is
validated when $r$ is colored by one of the $k$ colors $i_1,i_2,\cdots,i_k$.

If $r$ is not colored by one of the $k$ colors $i_1,i_2,\cdots, i_k$, then
$a_{i_\lambda}(T) =  \sum_f a_{i_\lambda}(T_f)$ for $1 \le \lambda \le k$. So
\begin{displaymath}
    a_{i_1}(T)+ \cdots + a_{i_k}(T) = \sum_f (a_{i_1}(T_f)+ \cdots + a_{i_k}(T_f)).
\end{displaymath}
The induction hypothesis on the height applies to each subtree $T_f$
because $h(T_f) < h(T)$.  Summing on all the subtrees,
\begin{displaymath}
    a_{i_1}(T)+ \cdots + a_{i_k}(T)  \le \sum_f (n_0(T_f)+\cdots + n_{k-1}(T_f)).
\end{displaymath}
As $\sum_f n_g(T_f) = n_g(T)$ for $g \le k-1$ because $k \le h(T)$,
Eq.~(\ref{eq:ineq_k}) is true in all cases and for all heights.

\section{Conclusion}

In this paper, we study a tree coloring problem described by
Guidon~\cite{Guidon-1,Guidon-2} based on an analogy with a remote control
system of a rail yard, seen as a switch tree.  Initially Guidon only
described binary trees, but this kind of coloring can be generalized to all
the rooted trees, binary or not.

To color a given tree $T$ of height $h$, the minimum number of colors is
$h+1$: this is understandable by considering for example the canonical
coloring by height, in which the $n_i$ nodes with height $i$ are colored with
label $i+1$.

The heart of this paper is the study of the distribution of the nodes of
$T$ among colors, more precisely the constraints on $A=(a_1, a_2,\cdots)$
where $a_i$ is the number of nodes with color $i$.  We show that the
sequence $A$ must satisfy a set of inequalities
Eqs.~(\ref{eq:sum}--\ref{eq:ineq_k}), or equivalently
Eqs.~(\ref{eq:sum},\ref{eq:a1},\ref{eq:max_1}--\ref{eq:max_k}), which only
involve macroscopic quantities of the tree,
the sequence $(n_0, n_1, \cdots)$.

We explain that there are trees $T$ for which these conditions are not
sufficient, i.e. there are sequences $A$ which check the inequalities, but
which cannot match a coloring of $T$.  It is possible even for full binary
trees, when they have a big imbalance between the branches, see
Fig.~\ref{fig:tftg}.

Incidentally, thanks to the canonical coloring of $T$ by depth in which the
$d_i$ nodes with depth $i$ are colored with label $i + 1$, the inequalities
are valid for two sequences of macroscopic quantities, the numbers of nodes
by depth $(d_0, d_1,\cdots)$ and the number of nodes by height
$(n_0, n_1,\cdots)$.

\subsection*{Acknowledgments}

It is a pleasure to thank Y.~Guidon for a stimulating discussion and for
sharing Ref.~\cite{Guidon-2} before publication.
This research did not receive any specific grant from funding agencies in
the public, commercial, or not-for-profit sectors.


\begin{thebibliography}{9}

  \bibitem{Knuth}
    D.~E.~Knuth,
    \emph{The art of computer programming, volume 1 – fundamental algorithms,}
    third ed., Reading, Massachusetts, 1997.

  \bibitem{Guidon-1}
    Y.~Guidon,
    \emph{\`{A} la d\'{e}couverte des arbres binaires \`{a} commande
          \'{e}quilibr\'{e}e,}
    GNU/Linux Magazine France \textbf{215} 10 (2018).

  \bibitem{Guidon-2}
    Y.~Guidon,
    \emph{Quelques applications des arbres binaires \`{a} commande
           \'{e}quilibr\'{e}e,}
    GNU/Linux Magazine France \textbf{218} 16 (2018).

  \bibitem{Stanley-book}
    R.~P.~Stanley,
    \emph{Catalan numbers,}
    Cambridge University Press, New York, 2015.
    https://doi.org/10.1017/CBO9781139871495

  \bibitem{OEIS}
    N.~J.~A.~Sloane, editor,
    \emph{The online encyclopedia of integer sequences,}
    published electronically at https://oeis.org/

\end{thebibliography}
\end{document}